\documentclass[aps,showpacs,preprintnumbers,amsmath,amssymb]{revtex4}
\usepackage{graphicx}
\usepackage{dcolumn}
\usepackage{bm}
\usepackage{epsfig}

\begin{document}

\title{\Large{Energy Loss of Charm Quarks in the Quark-Gluon Plasma: 
Collisional vs Radiative }}

\author{Munshi G. Mustafa }

\affiliation{Theory Group, Saha Institute of Nuclear Physics, 1/AF Bidhan Nagar,
Kolkata 700 064, India}

\vspace{0.2in}

\begin{abstract}
Considering the collisional energy loss rates of heavy quarks 
from hard light parton interactions the total energy loss of 
a charm quark for a static medium has been computed. For the energy range
$E\sim (5-10)$ GeV of charm quark, it is found to be almost same order as 
that of radiative ones estimated to a first order opacity expansion. 
The collisional energy loss will become much more important
for lower energy charm quarks and this feature could be very interesting for 
phenomenology of hadrons spectra.  Using such collisional energy 
loss rates
we estimate the momentum loss distribution employing a Fokker-Planck 
equation and the total energy loss of a charm quark for an expanding 
quark-gluon plasma under conditions resembling the RHIC energies.
The fractional collisional energy loss is found to be suppressed by a factor
of $5$ as compared to static case and does not depend linearly
on the system size.
We also investigate the heavy to 
light hadrons $D/\pi$ ratio at moderately large ($5-10$) GeV/$c$ 
transverse momenta and comment on its enhancement.
\end{abstract}


\maketitle

\vspace{0.2in}

\section{Introduction}
\label{sec_intro}

In the initial stage of ultra-relativistic heavy-ion collisions energetic 
partons are produced from hard collisions between the partons of the nuclei.
Receiving a large transverse momentum, 
these partons will propagate through the 
fireball which might consist of a quark-gluon phase for a transitional period
of a few fm/$c$. These high-energy partons will manifest themselves as jets
leaving the fireball. Owing to the interaction of the hard partons with
the fireball these partons will lose energy. Hence jet quenching will result.
The amount of quenching might depend on the state of matter of the fireball, 
i.e., quark-gluon plasma (QGP) or a hot hadron gas, respectively. Therefore
jet quenching has been proposed as a possible signature for the QGP formation
\cite{Pluemer}. 
Indeed, first results from $Au+Au$ at RHIC have shown a suppression of 
high-$p_\bot$ hadron spectra spectra \cite{RHIC} which could possibly indicate 
the quenching of light quark and gluon 
jets~\cite{Gyulassy,Salgado,Wang,Baier,muel,munshi0}. On the other hand the 
data~\cite{Light} from light ion interactions $D+Au$ at 
RHIC indicate no evidence of suppression in high $p_\bot$ hadron spectra 
implying the absence of jet quenching as there is no formation of extended 
dense medium in the final state in such light ion interactions. However,
this information from the light ion interactions in turn lends a strong 
circumstantial support that the observed suppression in 
$Au+Au$ is due to the final state energy loss of jets in the dense QGP matter.

Hadrons containing heavy quarks are important probes of 
strongly interacting matter produced in heavy ion collisions and has 
also excited a considerable interest. 
Heavy quark pairs are usually produced at early on a time scale of 
$1/2M_C\approx 0.07$fm/c from the initial fusion of partons
(mostly from $gg \rightarrow c{\bar c}$, but also from $q{\bar q}\rightarrow
c{\bar c}$) and also from quark-gluon plasma (QGP), if the initial temperature
is high enough. There is no production at late times in the QGP and none in 
the hadronic matter. Thus, the total number of charm quarks get frozen very
early in the history of collision which make them a good candidate for a probe
of QGP, as one is then left with the task of determining the $p_\bot$
distribution, whose details may reflect the developments in the plasma. 
The momenta distribution of $c$ quarks are likely to be reflected in the 
corresponding quantities in $D$ mesons as the $c$ quarks should pick up a 
light quark, which are in great abundance and hadronize.
The first
PHENIX data~\cite{Heavy} from RHIC in $Au+Au$ collisions at 
$\sqrt s=130$AGeV on prompt single electron production are now
available, which gives an opportunity to have an experimental estimate of the
$p_\bot$ distribution of the heavy quarks. Within the admittedly large 
experimental error the data indicate the absence of a QCD medium effects. We
hope that the future experimental study will provide data with improved 
statistics and wider $p_\bot$ range, which could then help us to understand 
the effect of medium modifications of the heavy quarks spectra. 

In order to see the effect of medium modifications on the final states,
the energy loss of hard partons in the QGP has to be determined. There are 
two contributions to the energy loss of a parton in the QGP: one is caused by
elastic collisions among the partons in the QGP and the other by radiation
of the decelerated color charge, {\it i.e.}, bremsstrahlung of gluons. 
The energy loss rates due to collisional scatterings among partons 
were estimated 
extensively~\cite{Bjorken,svet,Thoma1,Mrowczynski,Koike,Abhee,munshi,Hendrik} 
in the literature. 
Using the Hard-Thermal-Loop (HTL) resummed perturbative QCD at finite 
temperature \cite{BP}, the collisional energy loss of a heavy quark 
could be derived in a systematic 
way~\cite{Braaten1,Braaten2,Vija,Romatschke,Moore}.
It was also shown~\cite{svet,munshi,Hendrik,Moore} that the drag force can be related to the 
elastic scatterings among partons in a formulation based on the Fokker Planck 
equation  which is equivalent to the treatment of HTL 
approximations~\cite{Braaten2}. From these results also an estimate for the 
collisional energy loss of energetic gluons and light quarks
could be derived \cite{Thoma2}, which
was rederived later using the Leontovich relation \cite{Thoma3,Thoma4}.  

The energy loss due to multiple gluon radiation (bremsstrahlung)
was estimated and shown to be the dominant process. For a review 
on the radiative energy loss see Ref. \cite{BSZ}. Recently, it has also 
been shown~\cite{WW} that for a moderate value of the parton energy there is a
net reduction in the parton energy loss induced by multiple scattering due
to a partial cancellation between stimulated emission and thermal
absorption. This can cause a reduction of the light hadrons quenching 
factor as was
first anticipated in Ref.~\cite{munshi0}, though the most of the earlier
studies insisted that the collisional energy loss is insufficient to describe 
the medium modification of hadronic spectra.  
These studies, however, were limited to the case of
massless energetic quarks and gluons. 

The first estimate of heavy quark radiative energy loss was found to 
dominate~\cite{munshi1} the average energy loss rate and subsequently 
the charmed hadron~\cite{Vogt} and the dilepton~\cite{Shuryak} spectra had 
a strong dependence on the heavy quark radiative energy loss. 
Most recent studies of the medium modifications of the
charm quark spectrum have computed by emphasizing only the energy loss 
of heavy quarks by gluon 
bremsstrahlung~\cite{Dokshitzer,Djordjevic,Djordjevic1,Armesto}.
In Ref.~\cite{Dokshitzer} it was shown that the appearance of kinematical
dead cone effect due to the finite mass of the heavy quarks leads to a 
large reduction in radiative energy loss and affects significantly the
estimation of the quenching of charm quarks and $D/\pi$ ratio. Also 
in Refs.~\cite{Djordjevic,Djordjevic1,Armesto} a surprising degree of 
reduction in radiative energy loss for heavy quarks was obtained by 
taking into account the opacity expansion with and without 
the Ter-Mikayelian (TM) effect. 
In the framework in which all these calculations have been 
performed the heavy quarks are possibly not 
ultra-relativistic ($\gamma v \sim 1$)~\cite{svet,munshi,Hendrik,Moore} 
for much of the
measured momentum range and in this case it is far from clear that radiative
energy loss dominates over collisional case.

Rest of the paper is organized as follows: in Sec.~\ref{sec_eloss} 
the collisional energy loss for charm quarks is compared with 
the radiative ones computed in Refs.~\cite{Djordjevic,Djordjevic1}. 
The quenching of hadron spectra in a medium is briefly reviewed in
Sec.~\ref{sec_quen}. The charm quark in a thermally evolving plasma is modeled
in Sec.~\ref{sec_charm_evo} by an expanding fireball created in relativistic
heavy-ion collisions: we first obtain Fokker-Planck equation for a Brownian
particle from a generic kinetic equation (Sec.~\ref{sec_fp});   
we, next, compute analytically the momentum loss distribution for charm quarks 
using the collisional energy loss rates through the transport 
coefficients 
based on elastic perturbative cross section implemented into a Fokker-Planck 
equation (Sec.~\ref{sec_fp_evo}); the total energy loss for charm quark is
obtained for an expanding plasma (Sec.~\ref{sec_eloss_evo}); and then
the quenching of hadron spectra and $D/\pi$ is obtained for RHIC energies
(Sec.~\ref{sec_quench_evo}). We conclude in Sec.~\ref{sec_conl} with a brief 
discussion.


\section{Heavy Quarks in a Static Quark-Gluon Plasma}
\label{sec_eloss}
At leading order in strong coupling constant, $\alpha_s$, the energy loss
of a heavy quark comes from elastic scattering from thermal light 
quarks and gluons.  The energy loss rate of heavy quarks in the QGP
due to elastic collisions was estimated in Ref.~\cite{Braaten2}. In the 
domain $E<<M^2/T$, it reads
\begin{eqnarray} 
-\, \frac{d E}{dL}\, &=& \, \frac{8\pi\alpha_sT^2}{3} 
\left (1+\frac{n_f}{6}\right ) \left [ \frac{1}{v} - \frac{1-v^2}{2v^2}
\ln \left (\frac{1+v}{1-v}\right )\right ] \, 
\ln\left [ 2^{\frac{n_f}{6+n_f}} B(v) \frac{ET}{m_g M}\right ]  \, 
\label{rate1}
\end{eqnarray} 
whereas for $E>>M^2/T$, it is
\begin{eqnarray} 
-\, \frac{d E}{dL}\, &=& \, \frac{8\pi\alpha_sT^2}{3} 
\left (1+\frac{n_f}{6}\right ) \, 
\ln\left [ 2^{\frac{n_f}{2(6+n_f)}} 0.92 \frac{\sqrt{ET}}{m_g }\right ]  \, 
\label{rate2}
\end{eqnarray} 

where $n_f$ is the number of quark flavors, $\alpha_s$ is the strong coupling
constant, $m_g= \sqrt {(1+n_f/6)gT/3}$ is the thermal gluon mass, 
$E$ is the energy and $M$ is the mass of the heavy quarks. $B(v)$ is a smooth
velocity function, which can be taken approximately as $0.7$.
Following (\ref{rate1},\ref{rate2}), one can now estimate the static energy loss
for heavy quarks at the energies (temperatures) of interest.

\begin{figure}
\begin{minipage}[h]{0.48\textwidth}
\centering{\includegraphics[height=1.2\textwidth,width=1.0\textwidth]{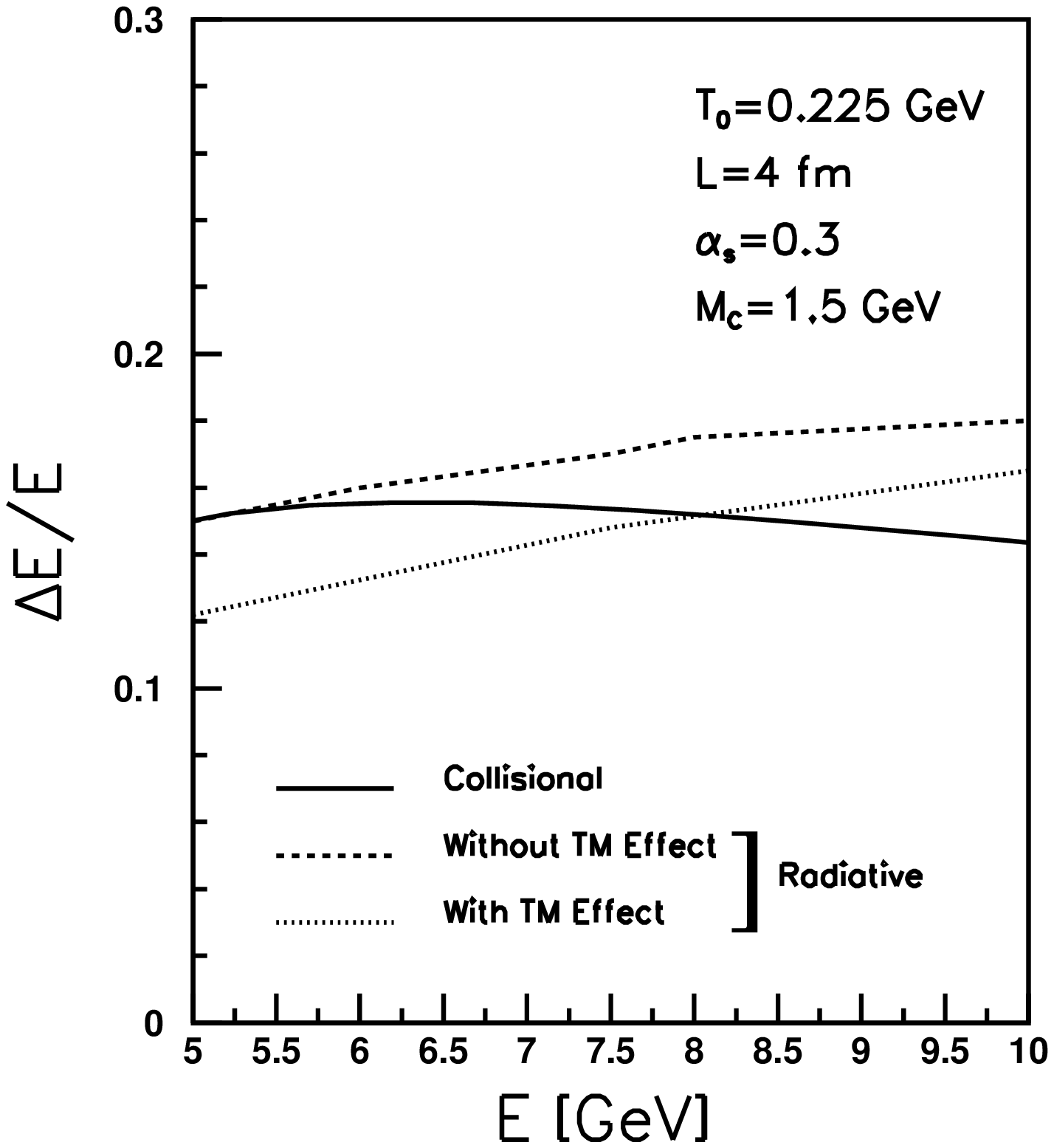}}
\end{minipage}
\begin{minipage}[h]{0.48\textwidth}
\centering{\includegraphics[height=1.2\textwidth,width=1.0\textwidth]{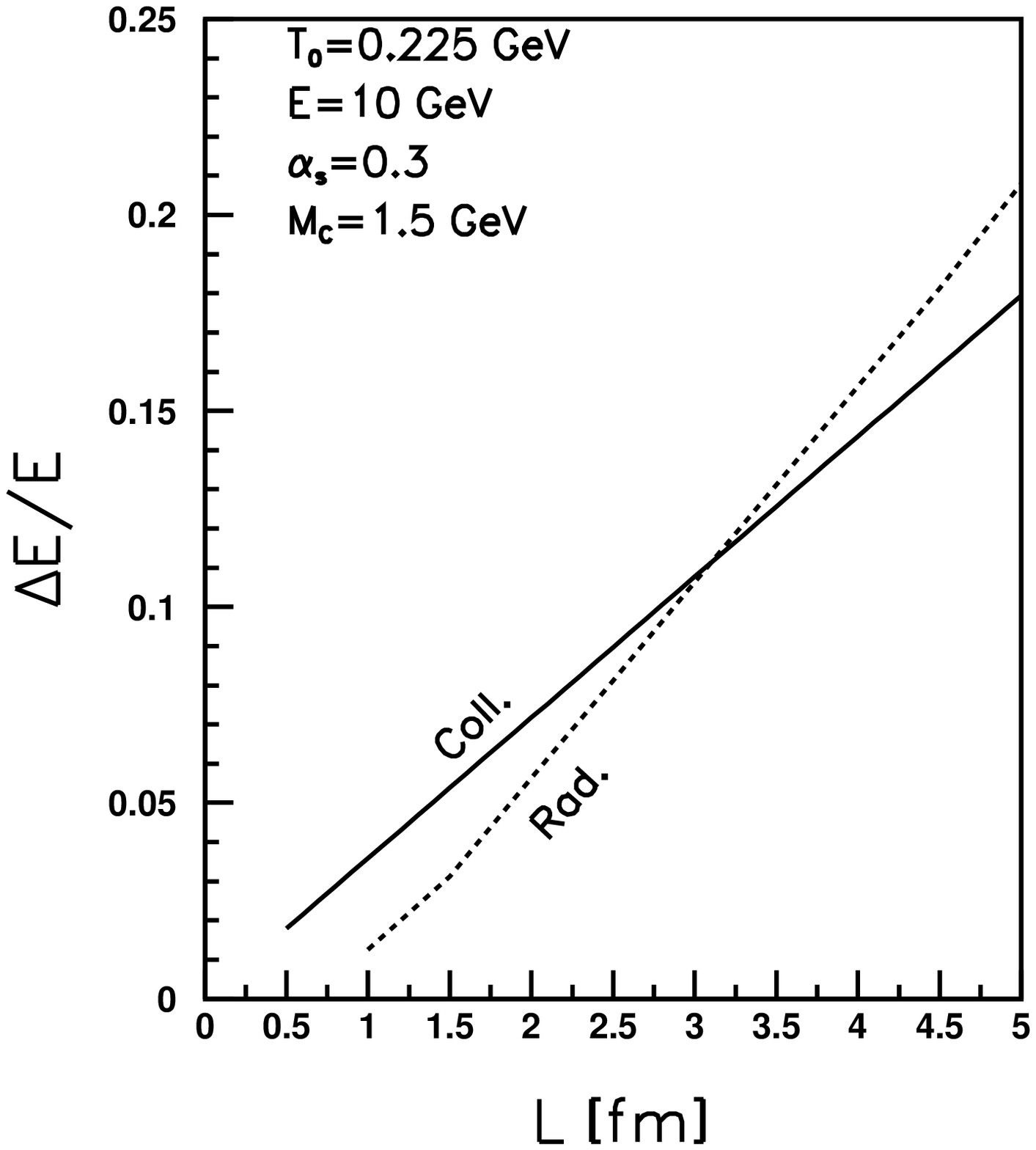}}
\end{minipage}
\vspace{-0.5in}
\caption{Left panel: The scaled static energy loss of a charm quark 
$\Delta E/E$ as a function of energy $E$ for a given length of the plasma,
$L=4$ fm. The collisional one is represented by solid line with plasma 
parameters for RHIC energy. The radiative energy losses according to
Ref.~\cite{Djordjevic1} are also plotted in 
first order opacity expansion with (dotted) and without (dashed)
the Ter-Mikayelian (TM) effect at a plasma length, $L=4$ fm and 
a fixed Debye screening mass, $\mu=0.5$ GeV (see text for details).
Right panel: The effective shift of the scaled
collisional (solid curve) and radiative (dashed) energy loss  
$\Delta E/E$ as a function
of distance $L$ for a charm quark of energy $E=10$ GeV.} 
\label{fig_static}
\end{figure}

On the other hand, heavy quarks medium induced radiative energy 
loss~\cite{Djordjevic,Djordjevic1} to all orders in 
opacity expansion, $L/\lambda_g$ ($L$ is the length of 
the plasma, $\lambda_g$ is the mean free path of the gluon), 
has been derived by generalizing the massless case~\cite{opacity} to
heavy quarks with mass in a QCD plasma with a gluon
dispersion characterized by an asymptotic plasmon mass. This also provides
the estimate of the influence of a plasma frequency cut-off 
on a gluon radiation (Ter-Mikayelian effect) 
and thus shields the collinear singularities
($k_\bot\rightarrow 0$) those arise due to massless quarks.

The medium induced radiative energy loss~\cite{Djordjevic1} for charm quark 
in first order opacity expansion, has been computed 
with a fixed
Debye screening mass, $\mu=0.5$ GeV with $\alpha_s=0.3$, and  a static plasma 
length, $L=4$ fm with $\lambda_g=1$ fm. 
The scaled energy loss was
found to obey a linear Bethe-Heitler like form, 
$\left.\frac{\Delta E}{E}\right |^{\rm{rad}}\propto L \sim CL$,  
where $C$ is constant of proportionality per unit length. 
The differential energy loss follows as 
$\left.\frac{d(\Delta E)}{dL}\right |^{\rm{rad}} \sim C E$. 
Now, $C$ can be estimated from right panel of Fig.~\ref{fig_static} 
(also from Fig.~2 of Ref.~\cite{Djordjevic1}), 
as $C\sim \frac{\Delta E/E}{L}\sim \frac{0.15}{4}$ fm$^{-1}$ 
at a plasma length $L=4$ fm,  $\mu=0.5$ GeV and $\alpha_s=0.3$. 
For a charm quark with energy $E=10$ GeV,
the differential radiative energy loss is estimated as 
$\left.\frac{d(\Delta E)}{dL}\right |^{\rm{rad}} \sim 0.375$ GeV/fm. 
The Debye screening mass is given as
$\mu = T\sqrt{4\pi \alpha_s(1+\frac{n_f}{6})}=2.2415T$, for two light
flavors, $n_f=2$ and $\alpha_s=0.3$. The Debye screening mass, $\mu=0.5$ GeV
corresponds to a temperature, $T=0.225$ GeV. 
With the  plasma parameters
corresponding to $\mu=0.5$ GeV, the differential collisional
energy loss  for a 10 GeV charm quark in a static medium 
can also be estimated from (\ref{rate1},\ref{rate2}) as  
$\left.\frac{d(\Delta E)}{dL}\right |^{\rm{coll}} \sim 0.36$ GeV/fm, 
and it is found to be of same order as that of radiative ones in
Ref.~\cite{Djordjevic1}. 

Now the total collisional energy loss can simply be evaluated from 
(\ref{rate1},\ref{rate2}). The scaled collisional (solid line)
and the radiative ones with (dotted) and without (dashed line) TM
effect of a
charm quark as a function energy $E$ are displayed in the left panel
of Fig.~\ref{fig_static} for a static plasma of length, $L=4$ fm,
with parameters $T=0.225$ GeV ($\mu=0.5$ GeV) and $\alpha_s=0.3$. 
As discussed earlier that the radiative energy loss is proportional
to $E$, resulting the scaled energy loss to be almost constant in $E$.
In the energy range, $E\sim(5-10)$ GeV, the scaled collisional 
energy loss is found to be almost similar as that of radiative
ones~\cite{Djordjevic1} but decreases with $E$ as the differential rates
in (\ref{rate1},\ref{rate2}) have dependence on a log factor involving $E$. 
Therefore, the collisional energy loss will become much more important for 
lower  energy range, and this feature itself will be quite interesting
for phenomenology of particle spectra.

In the right panel of Fig.~\ref{fig_static} we display the scaled 
effective energy loss of a charm quark
due to collisional (solid curve) and radiative~\cite{Djordjevic1} (dashed
curve) ones in  a static medium as a function of its thickness, 
$L$ for a given charm quark energy $E=10$ GeV. The thickness dependence of 
the scaled collisional energy loss for a given $E$ is linear 
like the radiative case~\cite{muel,Djordjevic1} whereas the earlier 
calculations~\cite{BSZ,opacity,Quad} show a quadratic form.  
This scaling clearly reflects a random walk in $E$ and $L$ as 
a fast parton moves in the medium~\cite{muel,Djordjevic1} with some 
interactions resulting in an energy gain and others in a loss of energy. 

In the energy range $(5-10)$ GeV, which is much of the experimentally 
measured range $(\gamma v \leq 4)$, the charm quark is not very 
ultra-relativistic and the collisions are found to be one of the the most  
dominant energy loss mechanisms.  
In the weak coupling limit bremsstrahlung~\cite{Moore} is the dominant 
energy loss mechanism if the charm quark is ultra-relativistic 
($\gamma v\geq 4$). 
Though the collisions have a different 
spectrum than radiation, the collisional rather than radiative energy loss 
should in principle determine the medium modifications of the final state 
hadron spectra. In the following 
we would study the suppression of heavy quark spectra.

\section{Quenching of hadron spectra}
\label{sec_quen}

We will follow the investigations by Baier et al. \cite{Baier} and 
M\"uller \cite{muel}, using the collisional instead of the radiative
parton energy loss. Following Ref.\cite{Baier} the $p_\bot$ distribution 
is given by the convolution of the transverse momentum distribution in 
elementary hadron-hadron collisions, evaluated at a shifted value 
$p_\bot+\epsilon$, with the probability distribution, $D(\epsilon )$,
in the energy $\epsilon$, lost by the partons to the medium by collisions,
as
\begin{eqnarray}
\frac{d\sigma^{\rm{med}}}{d^2p_\bot} &=& \int d\epsilon \, D(\epsilon) \,
\frac{d\sigma^{\rm{vac}}(p_\bot+\epsilon)}{d^2p_\bot}  
= \int d\epsilon \, D(\epsilon) \,
\frac{d\sigma^{\rm{vac}}}{d^2p_\bot}
+\int d\epsilon \, D(\epsilon) \, \epsilon \,
\frac{d}{dp_\bot} \frac{d\sigma^{\rm{vac}}}{d^2p_\bot}
+\cdots \cdots \nonumber \\
&=& \frac{d\sigma^{\rm{vac}}}{d^2p_\bot} + \Delta E \cdot 
\frac{d}{dp_\bot} \frac{d\sigma^{\rm{vac}}}{d^2p_\bot}
=  \frac{d\sigma^{\rm{vac}}(p_\bot+\Delta E)}{d^2p_\bot} 
= Q(p_\bot) \frac{d\sigma^{\rm{vac}}(p_\bot)}{d^2p_\bot}.
\label{rate}
\end{eqnarray}
Here $Q(p_\bot)$ is suppression factor due to the medium and 
the total energy loss by partons in the medium is
\begin{equation}
\Delta E = \int \epsilon \, D(\epsilon) \, d\epsilon \, . \label{eloss}
\end{equation}
We need to calculate the probability distribution, $D(\epsilon)$, that a 
parton loses the energy, $\epsilon$, due to the elastic collisions 
in the medium. This requires the evolution of the energy distribution of 
a particle undergoing Brownian motion, which will be obtained in
the  following Sec.~\ref{sec_charm_evo}.  

\section{Charm Quark in an Expanding Plasma}
\label{sec_charm_evo}
\subsection{Generic Kinetic Equation, Fokker-Planck Equation, 
Drag and Diffusion Coefficients}
\label{sec_fp}

The operative equation for the 
Brownian motion of a test particle can be obtained from the Boltzmann 
equation, whose covariant form can be written
as 
\begin{equation}
p^\mu\partial_\mu D({\mathbf {x,p,}}t) = C\{ D \} \, , \label{boltz}
\end{equation}
where $p^\mu(E_{\mathbf p},{\mathbf p})$ is the four momentum of the test 
particle, $C\{ D\}$ is the collision
term and $D({\mathbf {x,p,}}t)$ is the distribution due to the motion of the particle. 
If we assume a uniform plasma, the Boltzmann equation becomes
\begin{equation}
\frac{\partial D}{\partial t} = \frac{C\{ D  \}}{E}=
\left ( \frac{\partial D}{\partial t}\right )_{\rm{coll}} \, \, .
\label{unif}
\end{equation}
We intend to consider only the elastic collisions of the test parton 
with other partons in the background. The rate of collisions 
$w({{\mathbf p}}, {{\mathbf k}})$ is given by
\begin{equation}
w({{\mathbf p}}, {{\mathbf k}})= 
\sum_{j=q, {\bar q}, g} w^j(\mathbf{p,k}) \, , \label{coll}
\end{equation} 
where $w^j$ represents the collision rate of a test parton $i$ with other 
partons, $j$, in the plasma. The expression for $w^j$ can be written as
\begin{equation}
w^j({\mathbf{p,k}}) = \gamma_j \int \ \frac{d^3{\mathbf q}}{(2\pi)^3} 
D_j({\mathbf q}) v_{\rm{rel}} \sigma^j \, ,  \label{indiv}
\end{equation}
where $\gamma_j$ is the degeneracy factor, $v_{\rm{rel}}$ is the relative
velocity between the test particle and other participating partons $j$ from
the background, $D_j$ is the phase space density for the species $j$ and
$\sigma^j$ is the associated cross section. Due to this scattering the 
momentum of the test particle changes from $\mathbf p$ to $\mathbf{p-k}$. 
Then the collision term on the right-hand side of (\ref{unif}) can be 
written as
\begin{equation}
\left (\frac{\partial D}{\partial t}\right )_{\rm{coll}} = \int 
\ d^3{\mathbf k} \, 
\left [ w(\mathbf{p+k,k})D(\mathbf{p+k}) -w(\mathbf{p,k})D(\mathbf{p})
\right ] \, \, . \label{transi}
\end{equation}
where the collision term has two contributions. The first one is the
gain term where 
the transition rate $w(\mathbf{p+k,k})$ represents the rate that a particle with
momentum $\mathbf{p+k}$ loses momentum $\mathbf k$ due to the reaction with
the medium. The second term represents the loss due to the scattering of
a particle with momentum $\mathbf p$.

Now under the Landau approximation, i.e., most of the quark and gluon 
scattering is soft which implies that the function $w(\mathbf{p,k})$ is 
sharply peaked at $p\approx k$, one can expand the first term on
the right-hand side of (\ref{transi}) by a Taylor series as
\begin{equation}
w(\mathbf{p+k,k})D(\mathbf{p+k}) \approx w(\mathbf{p,k})D(\mathbf{p})+ 
{\mathbf k}\cdot \frac{\partial}
{\partial {\mathbf p}}(wD)+ \frac{1}{2}k_ik_j \frac{\partial^2}{\partial
p_i \partial p_j} (wD)+ \cdots \cdots \, . \label{taylor}
\end{equation}
Combining (\ref{unif}),(\ref{transi}) and (\ref{taylor}), one obtains
a generic kinetic equation of the form
\begin{equation}
\frac{\partial D}{\partial t} = \frac{\partial}{\partial p_i} 
\left [ {\cal T}_{1i}(\mathbf{p}) D \right ] + \frac{\partial^2}
{\partial p_i\partial p_j}
\left [ {\cal B}_{ij}(\mathbf{ p}) D \right ]\, \, , \label{ld}
\end{equation}
where the transport coefficients for momentum dispersion are given as
\begin{eqnarray}
{\cal T}_{1i}({{\mathbf p}}) &=& \int d^3{\mathbf k} \, 
w({{\mathbf p}}, {{\mathbf k}})\, k_i = 
 \int d^3{\mathbf k} \, w({{\mathbf p}}, {{\mathbf k}}) \, (p-p^\prime )_i 
 \, , \label{kern1} \\
{\cal B}_{ij}({\mathbf p}) &=&\frac{1}{2} \int d^3{\mathbf k}\, 
 w({{\mathbf p}}, {{\mathbf k}}) \, k_ik_j  
= \frac{1}{2} \int d^3{\mathbf k}\, w({{\mathbf p}},{{\mathbf k}})\, 
(p-p^\prime )_i(p-p^\prime )_j 
 \, \, . \label{kern2}
\end{eqnarray} 
These transport
coefficients in (\ref{kern1},\ref{kern2}) depend on the distribution function,
$D$, through the transition probability, $w(\mathbf{p,k})$ in (\ref{indiv}), 
 and can have
different values depending upon the problem.
The kinetic equation in (\ref{ld}) is the well known Landau 
equation~\cite{bales},  a non-linear integro-differential
equation, which describes, in general, collision 
processes between two particles. It should therefore depend, in a 
generic sense, on the states of two participating
particles in the collision process and hence on the product of two 
distribution functions making it non-linear in $D$. Therefore, it
requires to be solved in a self-consistent way, which is indeed a
non-trivial task. 

However, the problem can be simplified~\cite{bales} 
if one considers a large amount of weakly coupled particles in thermal
equilibrium at a temperature, $T$, constituting the heat bath in background
and due to the fluctuation there can be some non-thermal but homogeneously 
distributed particles constituting foreground. It is assumed that the overall 
equilibrium of the
bath will not be disturbed by the presence of such a few non-thermal particles.
Because of their scarcity, one can also assume that these non-thermal particles
will not interact among themselves but only with particles of the thermal 
bath in the background. This requires to replace the phase space distribution 
functions of the collision partners from heat bath appearing in (\ref{indiv})
by time independent or thermal distribution, $f_j(q)$. This will reduce 
the generic Landau kinetic equation (\ref{ld}), a non-linear 
integro-differential 
equation, to Fokker-Planck (FP) equation, a linear differential equation 
for the Brownian motion of the non-thermal particles in the foreground.

Now, one can write the transport coefficients in (\ref{kern1},\ref{kern2}) for 
such a FP equation in terms of the two body matrix elements, ${\cal M}$,
between a foreground and a background particles~\cite{svet,munshi}:
\begin{eqnarray}
{\cal T}_{1i}^{\rm{FP}}({\mathbf p}) &=& \frac{1}{2E_{\mathbf p}}\, 
\int \frac{d^3{\mathbf q}}{(2\pi)^3 2E_{\mathbf q}} \,
\int \frac{d^3{\mathbf {q^\prime}}}{(2\pi)^3 2E_{\mathbf {q^\prime}}}\,
\int \frac{d^3{\mathbf {p^\prime}}}{(2\pi)^3 2E_{\mathbf{p^\prime}}}\,
\frac{1}{\gamma_c} \sum |{\cal M}|^2 \, (2\pi)^4 \, 
\delta^4 (p+q-p^\prime-q^\prime) \nonumber \\
&& \times  
\left [p_i-p_i^\prime\right ] f({\mathbf q}){\tilde f}({\mathbf q}) 
\equiv \langle \langle ({p-p^\prime})_i\rangle\rangle \, \, , 
\label{transcolla} \\
{\cal B}_{ij}^{\rm{FP}}(\mathbf p) &=& \frac{1}{2} \langle\langle 
({p-p^\prime})_i ({p-p^\prime})_j \rangle\rangle  \, \, .
\label{transcollb} 
\end{eqnarray}
In our case, the incoming particle is a heavy quark which is different
from the background. So, $\mathbf p \, (\mathbf{p^\prime})$ and 
$\mathbf q \, (\mathbf{q^\prime})$ represent the momenta of the 
incoming (outgoing) charm- and background light-quark/gluon, respectively. 
For each background species, there is a similar additive contribution to
the collisional integral in (\ref{transcolla}). $\gamma_c$ is the spin and
color degeneracy factor of the foreground particle arising due to initial 
reaction channels.
$f(\mathbf q)$ is the particle distribution of the thermal background,
and 
$\tilde f(\mathbf q)= [1\pm f(\mathbf q)]$ corresponds a Bose enhancement/Pauli
suppression factor for scattered background particles, as appropriate. Because
of this thermal $f({\mathbf q})$ the contents of 
(\ref{transcolla},\ref{transcollb}) are different
from (\ref{kern1},\ref{kern2}) and the charm quark in the foreground 
of a weakly 
coupled system is driven by a Brownian motion 
mechanism~\cite{svet,munshi,Hendrik,Moore,somenath,rafelski,jane}.

We are now set to study the momentum distribution of a charm quark undergoing
Brownian motion and its relation with the transport coefficients. In absence
of vectors other than $\mathbf p$ the values of ${\cal T}_{1i}^{\rm{FP}}$ and
${\cal B}_{ij}^{\rm{FP}}$, which depend functionally on $\mathbf p$ and the 
background temperature $T$, must be of the form in  Langevin 
theory~\cite{svet,munshi,bales}:
\begin{eqnarray}
{\cal T}_{1i}^{\rm{FP}}({\mathbf p},T)&=&p_i{\cal A}(p^2,T) \, \, , 
\label{decomp1} \\
{\cal B}_{ij}^{\rm{FP}}({\mathbf p},T) &=& 
\left (\delta_{ij}-\frac{p_ip_j}{p^2}\right ){\cal B}_0(p^2,T) \,
+ \, \frac{p_ip_j}{p^2}{\cal B}_1(p^2,T) \, \, , \label{decomp2}
\end{eqnarray}
where, $p^2=p_i^2$, summation convention is always implied.
$\cal A$ is the drag, ${\cal B}_0$ is the transverse diffusion and ${\cal B}_1$
is the longitudinal diffusion coefficients. In terms of microscopic reaction
amplitudes these functions are obtained~\cite{svet,munshi} as
\begin{eqnarray}
{\cal A}(p^2,T)&=& \langle\langle 1 \rangle\rangle \, - \, 
\frac{\langle\langle \mathbf{p\cdot p^\prime}\rangle\rangle}{p^2}, 
\label{drag} \\
{\cal B}_0(p^2,T)&=& \frac{1}{4}\left[ 
\langle\langle {p^\prime}^2 \rangle\rangle
\, - \, \frac{\langle\langle(\mathbf{p\cdot p^\prime})^2\rangle\rangle}{p^2}
\right ] \, \, , \label{diffus0} \\
{\cal B}_1(p^2,T)&=& \frac{1}{2}\left [
 \frac{\langle\langle(\mathbf{p\cdot p^\prime})^2\rangle\rangle}{p^2}
\, - \, 2 \langle\langle \mathbf{p\cdot p^\prime}\rangle\rangle
\, + \, p^2 \langle\langle 1 \rangle\rangle  
\right ] \, \, . \label{diffus1}
\end{eqnarray} 
 The averaging, $\langle\langle\cdots\rangle\rangle$, defined in 
(\ref{transcolla}) 
can further be simplified~\cite{munshi}, by solving the kinematics in the 
centre-of-mass frame of the colliding particles, as
\begin{eqnarray}
\langle\langle F (\mathbf {p^\prime}) \rangle \rangle &=& 
\frac{1}{512\pi^4\gamma_c} \frac{1}{E_{\mathbf p}}\int_0^\infty \frac{q^2}
{E_{\mathbf q}} dq \int_{-1}^{1} d\left (\cos \chi \right )
\frac{\sqrt{(s+M_C^2-m_{g(q)}^2)^2-4sM_C^2}}{s}f(E_{\mathbf q}) \int_{-1}^{1}
d\cos\theta_{\rm{c.m.}} \nonumber \\
&& \, \times \,  \sum |{\cal M}|^2 \int_0^{2\pi} d\phi_{\rm{c.m.}} 
e^{\beta E_{\mathbf {q^\prime}}}f(E_{\mathbf{q^\prime}}) \, 
F(\mathbf{p^\prime}) \, \, , \label{expect}
\end{eqnarray}
where $M_C$ is the mass of a charm quark, 
$s=(E_{\mathbf p}+E_{\mathbf q})^2-(\mathbf{p+q})^2$,  
$E_{\mathbf{q^\prime}}=E_{\mathbf p}+E_{\mathbf q}-E_{\mathbf{p^\prime}}$ and 
$\mathbf{p^\prime}$ is a function of $\mathbf p$, $\mathbf q$ and 
$\theta_{\rm{c.m.}}$. ${\cal M}^2$ is the matrix 
elements~\cite{svet} for scattering processes $Qg, \, Qq, \, {\rm{and}}  
\, \, Q{\bar q}$, where $Q$ is a heavy quark and $g(q)$ is gluon(light quarks
with 2-flavors). 
The expression in (\ref{expect}) is larger by a 
factor of 2 than the ones originally derived in (3.6) of Ref.~\cite{svet}.
Apart from this we have also introduced the thermal masses of quarks ($m_q$),
gluons ($m_g$) and the quantum statistics, as appropriate.

The momentum and temperature dependence of the 
${\cal A}$, ${\cal B}_0$ and ${\cal B}_1$ are summarised in 
Figs. 1, 2 and 3 in Ref.~\cite{munshi} and we do not repeat them here and 
refer the reader to this work for details. The main finding is that these
coefficients are momentum independent upto $p=5$ GeV/$c$ and beyond this 
there is a weak momentum dependence. 
Note that the detailed studies of the dynamics of charm quark, 
as discussed in Refs.~\cite{svet}, may only depend on ${\cal A}$ and 
${\cal B}_0$, but perhaps not on ${\cal B}_1$ in the phenomenological 
relevant momentum range.
In the present calculation we are 
interested in the kinematical domain of $p=(5-10)$ GeV/$c$, for which
one needs to solve the FP equation considering the momentum dependence 
of drag and diffusion coefficients. This will require to solve the FP 
equation numerically. 

Instead, we assume the momentum independence of these 
coefficients in (\ref{drag},\ref{diffus0},\ref{diffus1}), which will
correspond to a scenario where a particle travels through an ideal
heat bath and undergoes linear damping (Rayleigh's particle). 
So, these transport coefficients are expected to be largely determined by the 
properties of the heat bath and not so much by the nature of the test
particle~\cite{bales}. 
This is also a fairly good approximation which we will
justify in the next section. Under this approximation, the transport
coefficients in (\ref{decomp1},\ref{decomp2}) become
\begin{eqnarray}
{\cal T}_{1i}^{\rm{FP}}&=& p_i {\cal A}
\label{iso1} \\
{\cal B}_{ij}^{\rm{FP}}&=& \delta_{ij}{\cal B}_0 \equiv 
\delta_{ij}{\cal T}_2^{\rm{FP}} \, \, , \label{iso2}
\end{eqnarray} 
where ${\cal B}_0({\mathbf p}\rightarrow 0,T)={\cal B}_1({\mathbf p}
\rightarrow 0, T) \equiv {\cal T}_2^{\rm{FP}}$.  
This could be viewed as a
course-grained picture in which the finer details of the collisions have been
averaged out over a large number of macroscopic situation 
(or over an ensemble). 
Then combining (\ref{ld}) and (\ref{iso1},\ref{iso2})
one can write the FP equation as
\begin{equation}
\frac{\partial D}{\partial t} = \frac{\partial}{\partial p_i} 
\left [ {\cal T}_{1i}^{\rm{FP}} D \right ] + {\cal T}_2^{\rm{FP}} 
\left ( \frac{\partial} {\partial {\mathbf p}}\right )^2
 D \, \, . \label{isofp}
\end{equation}
In the next Sec.~\ref{sec_fp_evo} we obtain the time evolution of 
FP equation in a thermally evolving QGP.

\subsection{Time Evolution of Fokker-Planck Equation, Drag and Diffusion Coefficients in an Expanding Plasma}
\label{sec_fp_evo}

We assume that the background partonic system has achieved thermal 
equilibrium when the momenta of the background partons become locally
isotropic. At the collider energies it has been estimated that 
$t_0=0.2-0.3$ fm/$c$. Beyond this point, the further expansion is 
assumed to be described by Bjorken scaling law~\cite{bjor} 
\begin{equation}
T(t)=t_0^{1/3} T_0/t^{1/3},  \label{bjscal} 
\end{equation}
where $T_0$ is the initial temperature at which background has attained
local thermal equilibrium.

We consider, for simplicity, the one dimensional problem,  for which
FP equation in (\ref{isofp}) reduces to 
\begin{equation}
\frac{\partial D}{\partial t} = \frac{\partial}{\partial p} 
\left [ {\cal T}_{1}^{\rm{FP}} D \right ] + {\cal T}_2^{\rm{FP}} 
\frac{\partial^2} {\partial {p}^2} D \, \, , \label{fp1}
\end{equation}
and as discussed in the previous Sec.~\ref{sec_fp} 
that 
the coupling between the Brownian particle and the bath
is weak, the quantities ${\cal T}_1^{\rm{FP}}$ and
${\cal T}_2^{\rm{FP}}$ can also be written 
using the Langevin formalism~\cite{bales} as
\begin{eqnarray}
{\cal T}_1^{\rm{FP}}(p)& = &\int dk \, w(p,k)\, k = 
\frac {\langle \delta p \rangle}
{\delta t} = \langle F \rangle = p {\cal A}
\label{1diso1} \\ 
{\cal T}_2^{\rm{FP}} &=&\frac{1}{2}  
\frac {\langle (\delta p )^2\rangle}
 {\delta t} \approx  T{\cal T}_1^{\rm{FP}} \, \, . \label{diff}
\end{eqnarray} 

Now the work done by the drag force, ${\cal T}_1^{\rm{FP}}$, acting on a test particle is
\begin{equation}
-dE = \langle F\rangle \cdot dL = {\cal T}_1^{\rm{FP}}
\cdot dL \, \, ,\label{work}
\end{equation}
which can be related to the energy loss~\cite{Braaten2,Thoma2} of a particle as
\begin{equation}
-\frac{dE}{dL} = {\cal T}_1^{\rm{FP}} = p\, {\cal A}  \, \, . \label{celoss}
\end{equation}
The drag coefficient is a very important quantity containing the
dynamics of elastic collisions and it has a weak momentum dependence. Then 
one can average out the drag coefficient as
\begin{equation}
\langle {\cal A}(p,T(t)) \rangle \, \equiv \, {\cal A}(T(t)) \, = \, 
\left \langle -\frac{1}{p}\, \, \frac{dE}{dL}
 \right \rangle 
\, \, , \label{average}
\end{equation}
implying that the dynamics is solely determined by the collisions 
in the heat bath and
independent of the initial momentum of the Brownian particle.

\begin{figure}
\begin{minipage}[h]{0.48\textwidth}
\centering{\includegraphics[height=1.2\textwidth,width=1.0\textwidth]{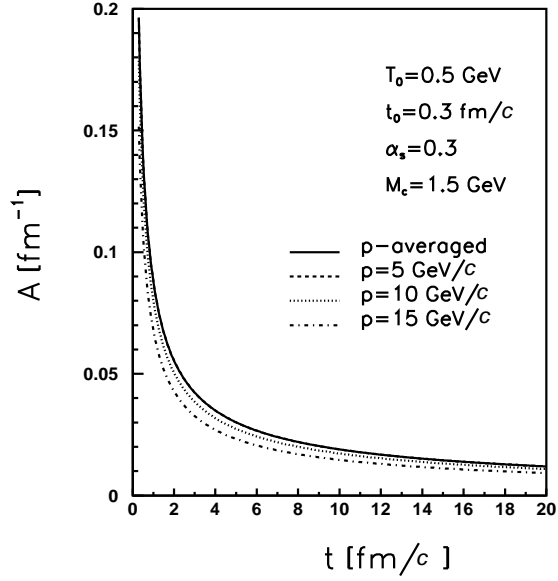}}
\end{minipage}
\vspace{-0.5in}
\caption{ The momentum averaged ($\langle{\cal A}(t)\rangle$) in 
(\ref{average}) and momentum dependence (${\cal A}(p,t)$) in (\ref{celoss})
of the drag coefficient of a charm 
quark in an expanding QGP with plasma parameters (see text) suitable for 
RHIC energy.}
\label{fig_drag}
\end{figure}

For averaging over the momentum the Boltzmann distribution and
the differential energy loss rates (\ref{rate1},\ref{rate2}) were 
used.
The time dependence of the drag coefficient comes from assuming
a temperature, $T(t)$, decreasing with time as the system expands, 
according to Bjorken scaling law~\cite{bjor} given in (\ref{bjscal}).
We consider the initial 
temperature $T_0=0.5$ GeV, initial time $t_0=0.3$ fm/$c$ and $\alpha_s=0.3$ 
of the plasma for RHIC energy.
In Fig.~\ref{fig_drag} the momentum averaged as well the momentum dependence of
drag coefficient of a charm quark in the 
QGP phase of the expanding fireball is shown as a function of the time. As
can be seen that the behavior of the momentum averaged drag coefficient 
(solid curve) is dominated by $T^2/p \sim t^{-1/3}$ according to the scaling 
law. Now, it can also be seen that upto $p=10$ GeV/$c$ there is no significant 
difference between momentum averaged, $\langle{\cal A}(T(t))\rangle$,
 and momentum dependence, ${\cal A}(p,T(t))$, of drag 
coefficient and it has only a weak $p$ dependence beyond $p= 10$ GeV/$c$. 
Since it decreases with moderately high values of $p$ ($\geq15$ GeV/$c$), 
the momentum averaged approximation of drag coefficient,
$\left \langle{\cal A}(T(t))\right \rangle$, would overestimate the actual 
${\cal A}(p,T(t))$ in
this high momentum range.  
In our phenomenological approach the momentum independence of drag 
coefficient
in (\ref{average}) is a good approximation upto a moderate value of 
momentum $p\leq 15$ GeV/$c$. 

\begin{figure}
\begin{minipage}[h]{0.48\textwidth}
\centering{\includegraphics[height=1.2\textwidth,width=1.0\textwidth]{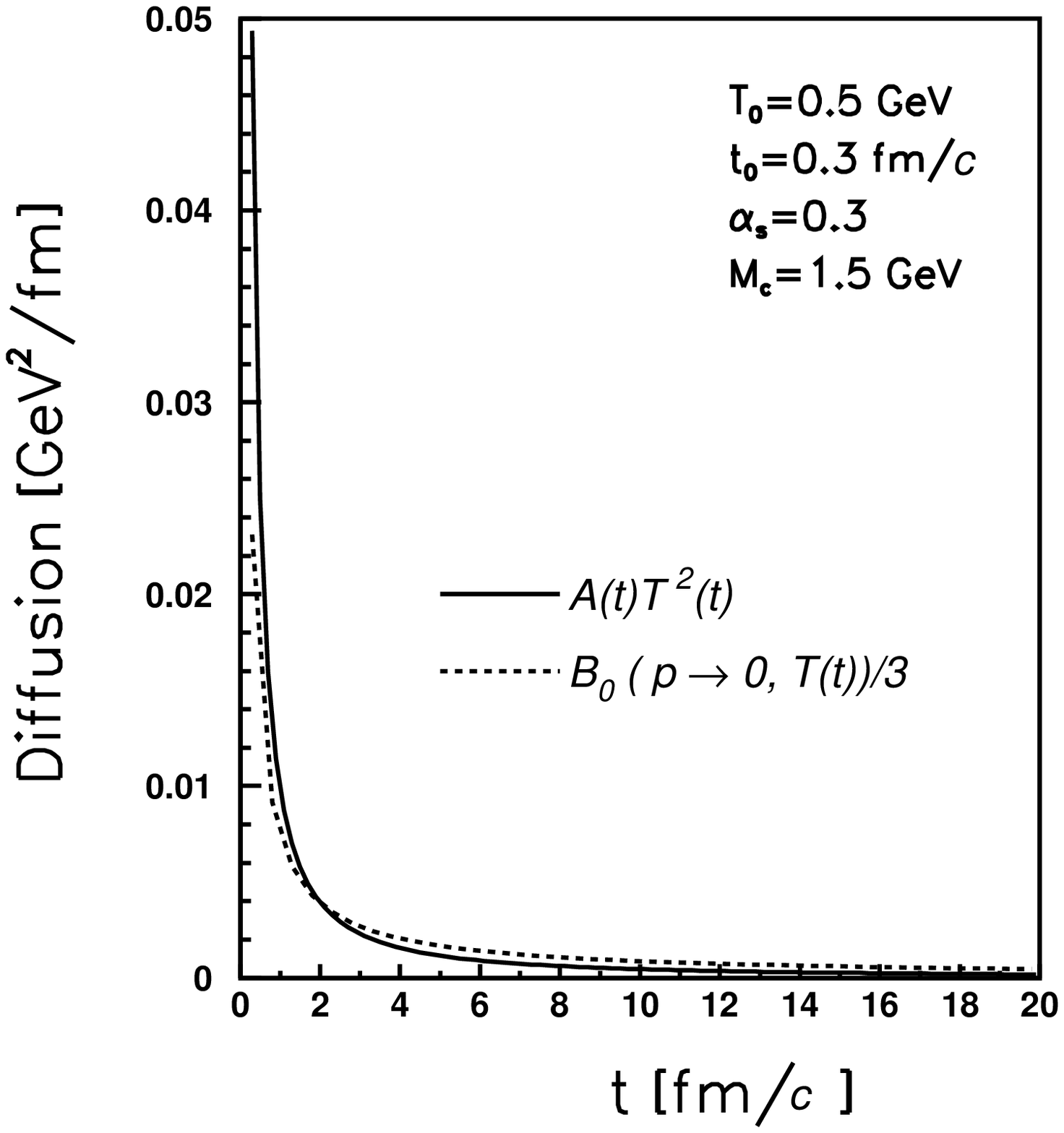}}
\end{minipage}
\begin{minipage}[h]{0.48\textwidth}
\centering{\includegraphics[height=1.2\textwidth,width=1.0\textwidth]{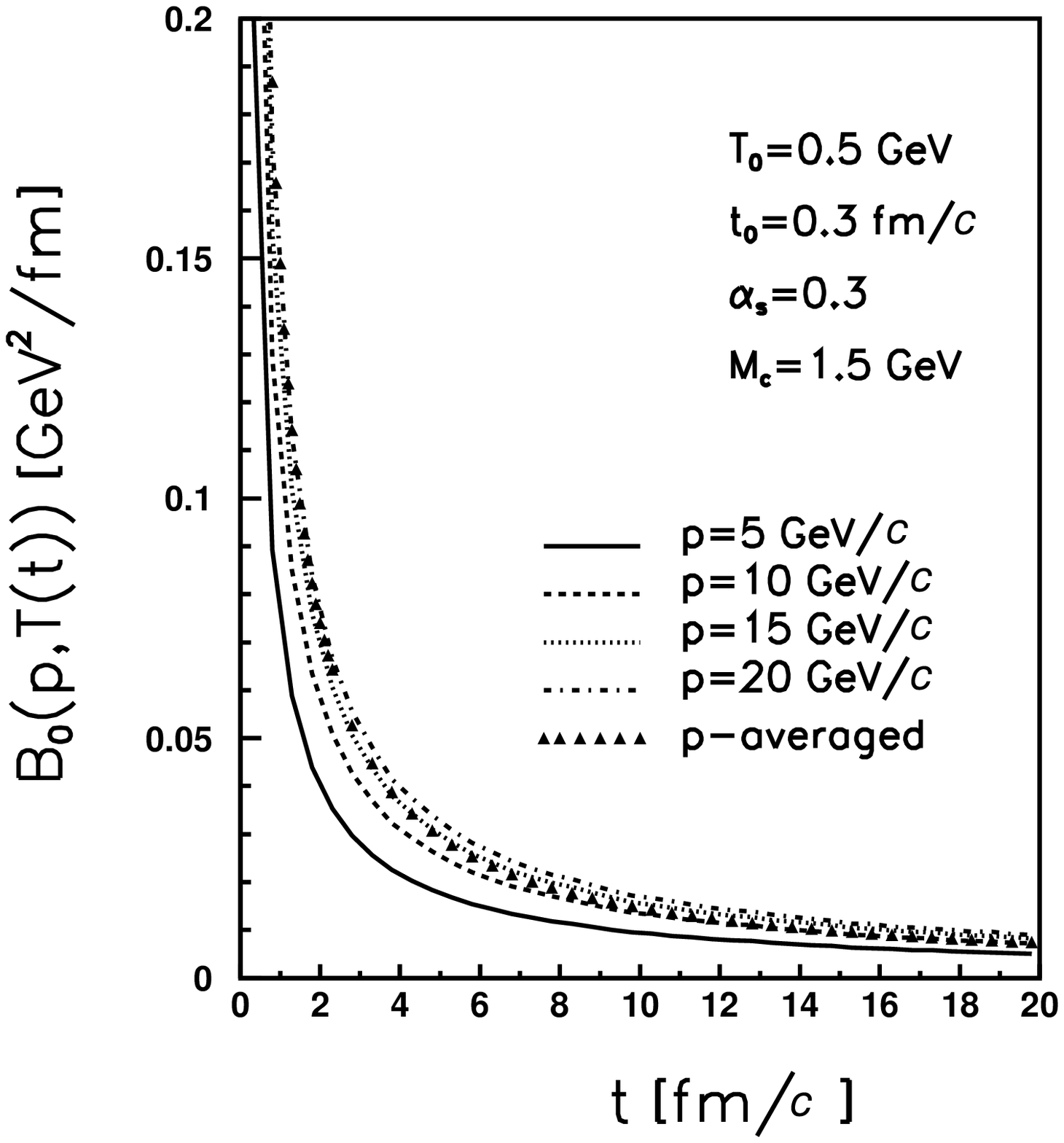}}
\end{minipage}
\vspace{-0.5in}
\caption{ Left panel: Comparison of diffusion coefficients given in
(\ref{diff1}) (solid line) and  zero momentum limit of ${\cal B}_0$ 
(dashed line) given in (\ref{diffus0}). \, \, 
Right panel: Comparison of momentum averaged $\langle {\cal T}_2^{\rm{FP}}
\rangle$ in (\ref{avgdiff}) and momentum
dependence of ${\cal B}_0$ in (\ref{diffus0}).}
\label{fig_diff}
\end{figure}

Now, combining (\ref{1diso1}) and (\ref{diff}) we can write the diffusion 
coefficient as 
\begin{equation}
{\cal T}_2^{\rm{FP}} = T \, {\cal T}_1^{\rm{FP}} = T \, {\cal A} \, p
\, \, . \label{1diso2a}
\end{equation}
Once the drag coefficient is averaged out using the properties of heat bath,
one can approximate $p$ by the temperature $T$ of the bath (as discussed 
earlier that it is independent of initial momentum of the Brownian particle)
and 
${\cal A}$ by its average value given in (\ref{average}). This leads 
\begin{equation}
{\cal T}_2^{\rm{FP}} = {\cal A}(T(t)) T^2(t) 
\, \, , \label{diff1} 
\end{equation}
which is also known as
the Einstein relation~\cite{bales} between drag and diffusion coefficients. 
In the left panel of Fig.~\ref{fig_diff} the diffusion coefficient obtained 
in (\ref{diff1}) represented by solid line 
is displayed. It is found to have agreed quite well with the momentum 
independent diffusion coefficient (dashed line), 
${\cal B}_0(p\rightarrow 0)$, 
in (\ref{diffus0}) with a
factor $1/3$ multiplied with it, because we consider the 1-dimensional
scenario. As evident that momentum independence of diffusion coefficient is 
also a fairly good approximation.

Alternatively, one can also calculate the ${\cal T}_2^{\rm{FP}}$ in 
(\ref{1diso2a}) by substituting ${\cal A}$ from (\ref{celoss}) and 
 averaging out the momentum dependence as
\begin{equation}
\langle {\cal T}_2^{\rm {FP}} \rangle = T \left\langle -\frac{dE}{dL} 
\right\rangle  
\, . \label{avgdiff}
\end{equation}
In the right panel of Fig.~\ref{fig_diff} the average diffusion coefficient
computed in (\ref{avgdiff}) (filled triangle) is displayed. The momentum 
dependence transverse diffusion coefficient, ${\cal B}_0(p^2,T(t))$, 
in (\ref{diffus0}) is also plotted for different momenta. 
It can be seen that there is weak momentum dependence in ${\cal B}_0$ 
in the momentum range $p=5-20$ GeV/$c$. The momentum averaged values are in
agreement with ${\cal B}_0$ for higher momenta $p\ge 10$ GeV/$c$ whereas it 
overestimates lower momenta $p< 10$ GeV/$c$. We will use both the 
approximations for diffusion coefficient in (\ref{diff1},\ref{avgdiff})
in our purpose to obtain the momentum distribution below.

Combining (\ref{fp1}) and (\ref{celoss}), we find
\begin{equation}
\frac{\partial D}{\partial t} = {\cal A} \frac{\partial }{\partial p} (p D)
+ {\cal D}_F\frac{\partial^2 D}{\partial p^2} \, \, , \label{landau} 
\end{equation}
where ${\cal D}_F$, in general, has been used as the diffusion coefficient 
corresponding to (\ref{diff1},\ref{avgdiff}), 
and ${\cal A}$ is the averaged drag coefficient in (\ref{average}).

Next we proceed with solving the above equation with the boundary condition
\begin{equation}
D(p,t)\, \, \, \stackrel{t\rightarrow t_0}{\longrightarrow}
\, \, \delta(p-p_0)\, \, . \label{boundc}
\end{equation}

The solution of (\ref{landau}) can be found by making a
Fourier transform of $D(p,t)$,
\begin{equation}
D(p,t)=\frac{1}{2\pi}\int_{-\infty}^{+\infty} \, {\tilde D}(x,t) \, 
\, e^{ipx}dx 
\, \, , \label{fourr}
\end{equation}
where the inverse transform is
\begin{equation}
{\tilde D}(x,t)=\int_{-\infty}^{+\infty} \,  D(p,t) \, \, e^{-ipx}dp 
\, \, . \label{invfourr}
\end{equation}
Under the Fourier transform the corresponding initial condition follows
from (\ref{boundc}) and (\ref{invfourr}) as
\begin{equation}
{\tilde D} (x_0,t=t_0)=e^{-ip_0 x_0} \, \, \label{bcfourr}
\end{equation}
where $x=x_0$ at $t=t_0$ is assumed. 

Replacing 
$p\rightarrow i\frac{\partial}{\partial x}$ and  
$\frac{\partial}{\partial p}\rightarrow ix$, the Fourier transform of
(\ref{landau}) becomes
\begin{equation}
\frac{\partial {\tilde D}}{\partial t} + {\cal A}x \frac{\partial {\tilde D}}
{\partial x} 
= -{\cal D}_F \, x^2 {\tilde D} \, . \label{landau4} 
\end{equation}
This is a first order partial differential equation which may be solved by
the method of characteristics~\cite{william}. The characteristic
equation corresponding to (\ref{landau4}) reads
\begin{equation}
\frac{\partial t}{1} \, = \, \frac{\partial x}{{\cal A} x}\, =\,
 -\frac{\partial {\tilde D}} {{\cal D}_Fx^2{\tilde D}} \, \, \, \, 
. \label{charac}
\end{equation}


Using the boundary condition in (\ref{bcfourr}) 
the solution of (\ref{landau4}) can
be obtained as
\begin{eqnarray}
D(p,L) &=& \frac{1}{\sqrt{\pi\, {\cal W}(L)}} \, \exp \left [ 
- \frac{\left (p-p_0\, e^{-\int^L_0{\cal A}(t') \, dt'} \right )^2}
{{\cal W}(L)} \right ] \, \, , \label{solfin} 
\end{eqnarray}
where ${\cal W}(L)$ is given  by
\begin {equation}
{\cal W}(L) = \left ({4\int_0^L 
{\cal D}_F(t')
\exp \left [ 2 \int^{t'} {\cal A}(t'')\, dt'' \right ]\, dt'}\right )
\left [{\exp \left (-2 \int_0^L {\cal A}(t')\, dt'\right )} \right ] \, \, ,
\label{gauswid1}
\end{equation}
which is the probability distribution in momentum space. 
Since the plasma 
expands with the passage of time, 
we have used the length of the plasma, $L$ as the maximum time limit 
for the relativistic case ($\gamma v\sim 1$). 
\begin{figure}
\begin{minipage}[h]{0.48\textwidth}
\centering{\includegraphics[height=1.2\textwidth,width=1.0\textwidth]{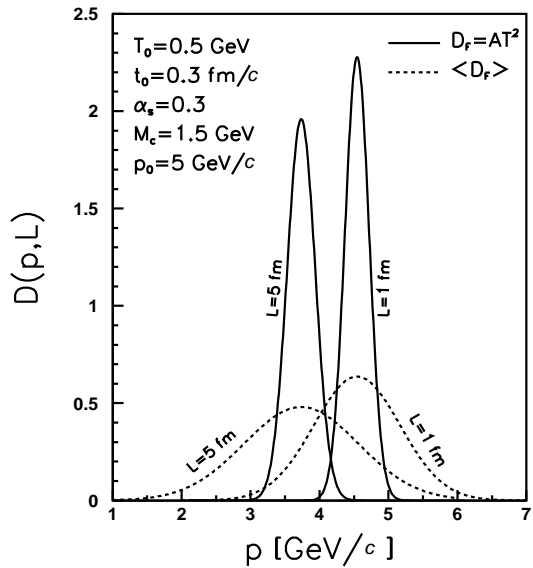}}
\end{minipage}
\vspace{-0.5in}
\caption{ 
The momentum-loss probability distribution, $D(p,L)$ 
of a charm quark as a function of momentum $p$ after plasma has expanded 
a distance $L$.} 
\label{fig_dist}
\end{figure}

In Fig.~\ref{fig_dist} we show the momentum loss probability distribution 
$D(p,L)$ given in (\ref{solfin}), of a charm quark with initial momentum,
$p_0=5$ GeV/$c$, as a function of momentum $p$. The solid lines represent
the distribution with the diffusion coefficient, ${\cal D}_F={\cal A}T^2$,
in (\ref{diff1}) whereas the dashed lines with 
$\langle {\cal D}_F\rangle$ in (\ref{avgdiff}). Both set of curves are
for two different expanded plasma lengths, $L=1$ fm and $5$ fm as indicated in 
Fig.~\ref{fig_dist}. In general, the physical mechanism reflected in this 
Fig.~\ref{fig_dist} can be understood at initial time ($t_0$) or length 
of the plasma a momentum distribution is sharply peaked at $p=p_0$,
according to (\ref{boundc}).  With passage of time (or distance traveled) 
the peak of the probability distribution is shifted towards smaller momentum,
as a result of drag force is acting on the momentum of the charm quark 
indicating its most probable momentum loss  due to elastic
collisions in the medium. Moreover, the peak broadens slowly as a
result of diffusion in momentum space, implying that a finite momentum 
dispersion sets in. As evident, with both the approximations in diffusion 
coefficient, only the momentum dispersion is affected while the 
peak positions remains unaltered, indicating that a drag force acting on
the mean momentum of a charm quark is same.
After plasma has expanded upto a length, $L=1$ fm the charm quark loses
$10\%$ of its momentum whereas it is 25$\%$ at a expanded length
 of $L=5$ fm.
In the next subsec.~\ref{sec_eloss_evo}, we will use this distribution
to compute the total energy loss of a charm quark for an expanding plasma.

\subsection{Energy-loss of a charm quark in an expanding plasma}
\label{sec_eloss_evo}

In the preceding subsec.~\ref{sec_fp_evo}
we obtain a momentum-loss distribution by solving the
time evolution of FP equation in a thermally
evolving plasma, which is modeled by an expanding fireball under conditions
resembling central {\it Au-Au} collisions at RHIC. 
The mean energy of a charm quark due to 
the elastic collisions in a expanding  medium can be estimated as 
\begin{equation}
\langle \, E \, \rangle \, = \, \int_0^\infty \, E \, D(p,L) \, dp \, \, . 
\label{meane} 
\end{equation}
The average energy loss due to elastic collisions in the medium is 
given by
\begin{eqnarray}
\Delta E\, = \langle \, \epsilon \, \rangle \, &=& \, 
E_0\, -\, \langle \, E \, \rangle  \,  
\, \, , \label{avgelos}
\end{eqnarray}
where $E=m_\bot \, = \sqrt{p_\bot^2 \, + m^2 }$ at the central
rapidity region, $y=0$. 


\begin{figure}
\begin{minipage}[h]{0.48\textwidth}
\centering{\includegraphics[height=1.2\textwidth,width=1.0\textwidth]{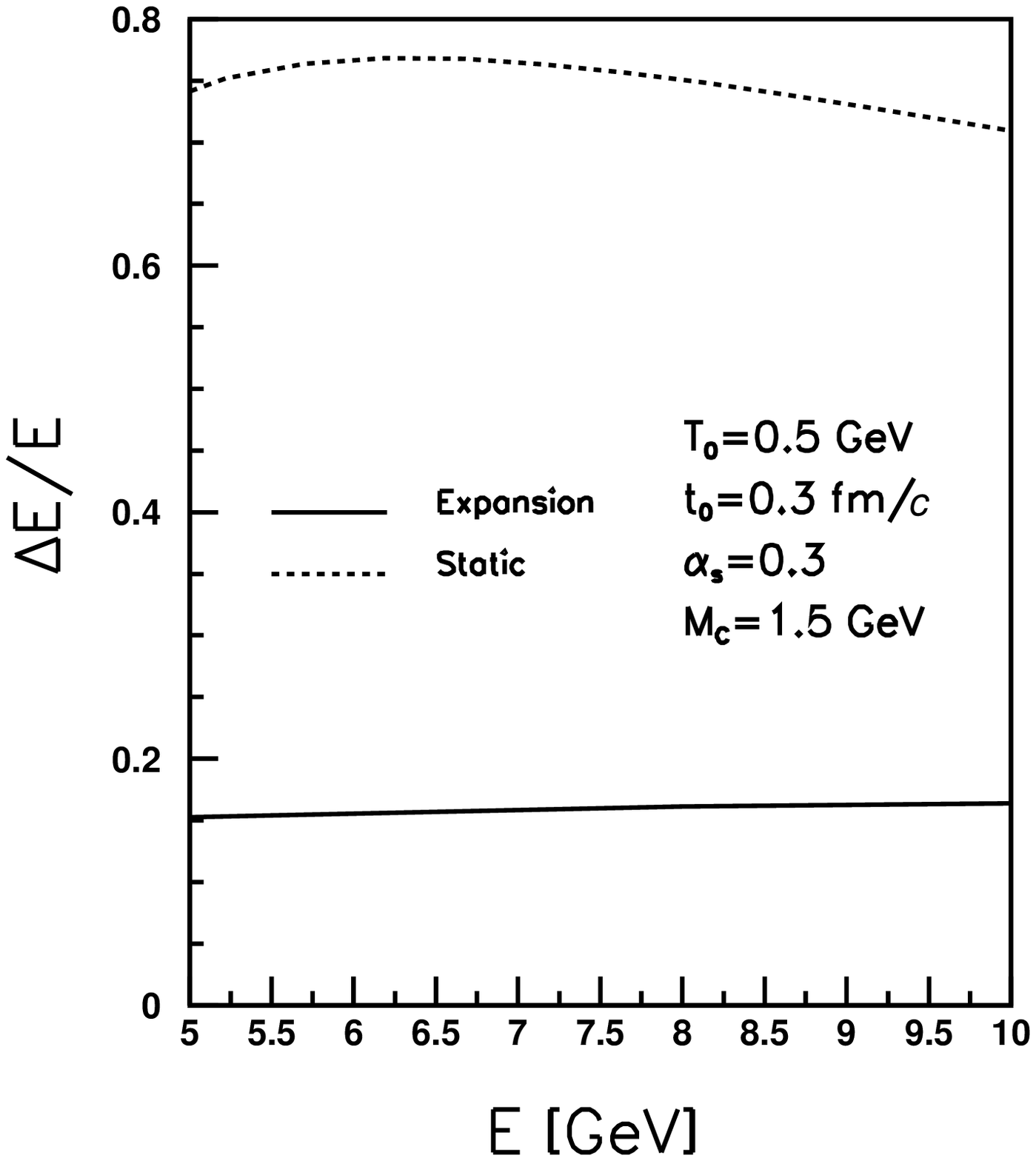}}
\end{minipage}
\begin{minipage}[h]{0.48\textwidth}
\centering{\includegraphics[height=1.2\textwidth,width=1.0\textwidth]{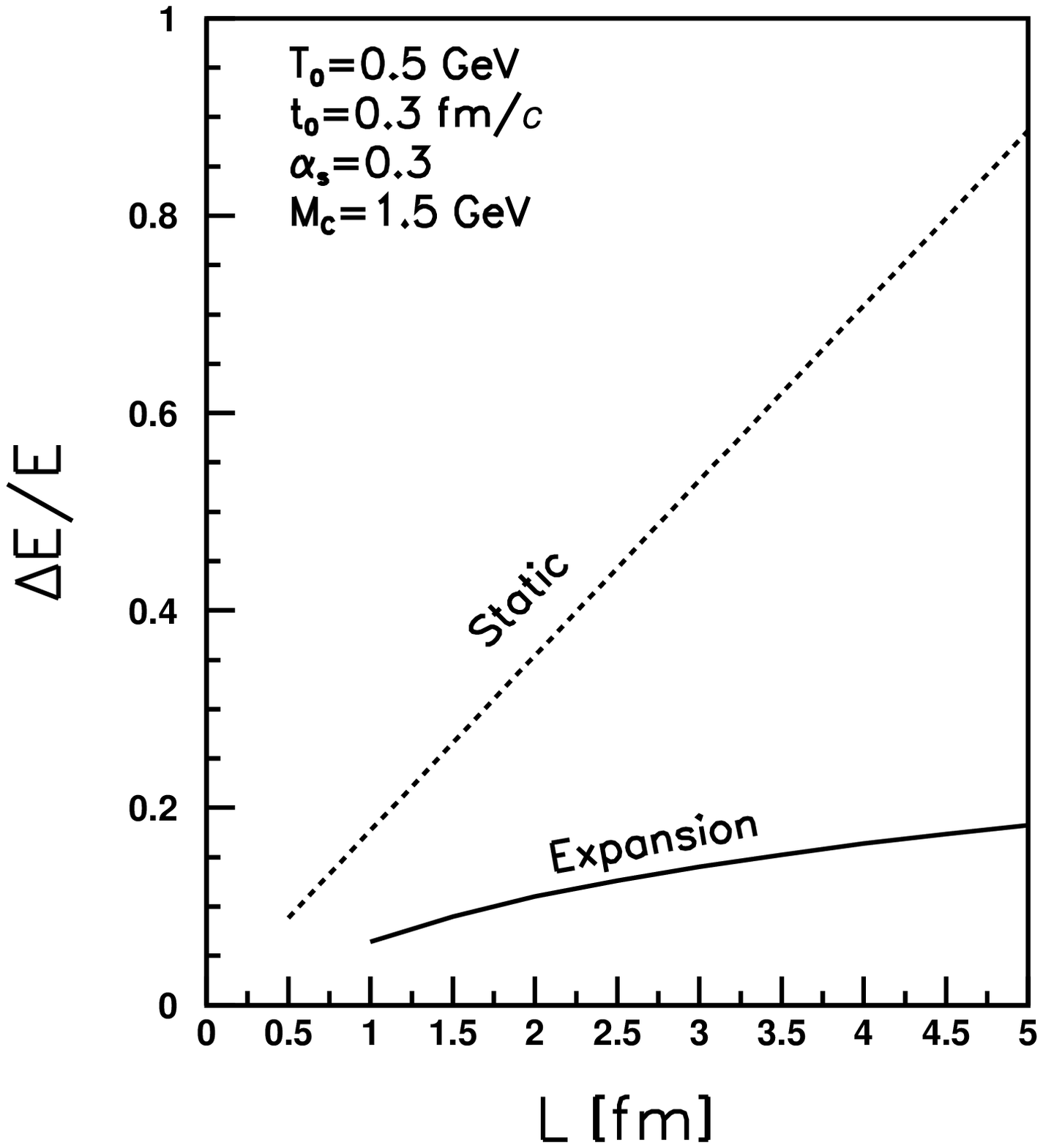}}
\end{minipage}
\vspace{-0.5in}
\caption{Left panel: 
The fractional collisional energy loss of a charm quark 
$\Delta E/E$ as a function of energy $E$ when plasma has expanded upto
$L=4$ fm (solid line) and for static plasma of length, $L=4$ fm (dashed
line). 
Right panel: Collisional  
$\Delta E/E$ as a function of length $L$ for a charm quark of $E=10$ GeV 
for expanding (solid line) and static plasma (dashed line).
} 
\label{fig_eloss_evo}
\end{figure}

Using the momentum loss distribution in (\ref{solfin}) the total energy loss of 
a charm quark has been computed in (\ref{avgelos}).
The numerical results for scaled collisional energy loss of a
charm quark as a function of energy $E$ in an expanding plasma (solid line) 
is shown in the left panel of Fig.~\ref{fig_eloss_evo} for the plasma 
parameters $T_0=0.5$ GeV, $t_0=0.3$ fm/$c$, 
$\alpha_s=0.3$ and expanded plasma length upto $L=4$ fm. 
In the energy range, $E \sim 5-10$ GeV,  the fractional collisional 
energy loss remains almost constant around a value $0.15$ and the reason for
which can be traced back to the momentum independence of the drag 
coefficient~\cite{svet,munshi} as discussed earlier. The corresponding 
scaled energy loss for a static plasma is shown by dashed line. Taking into 
account the expansion, the scaled energy loss is suppressed by 
a factor of $5$ as compared to static case. In the right panel of 
Fig~\ref{fig_eloss_evo} the scaled energy loss is plotted as a function
of length, $L$ for a given energy, $E=10$ GeV and found that it does not
depend linearly on the system size for the expanding case (solid line) as 
compared to static case (dashed line). It is also expected that the
similar suppression should also occur for radiative case due to the 
expansion~\cite{Wang1}. 

\subsection{Quenching of Hadron Spectra in an expanding plasma}
\label{sec_quench_evo}

We assume that the geometry is described by a cylinder of radius $R$,
as in the Boost invariant Bjorken model~\cite{bjor} of nuclear collisions,
and the parton moves in the transverse plane in the local rest frame.
Then a parton created at a point $\vec{\mathbf r}$ with an angle $\phi$
in the transverse direction will travel a distance \cite{muel}

\begin{equation}
L(\phi)= (R^2\, - \, r^2 \, \sin^2\phi \,)^{1/2}\, - \, r \, \cos \phi \, \,
\, , \label{trad}
\end{equation}
where $\cos \phi\, = \,{\hat {\vec {\mathbf v}}} \, \cdot \, 
{\hat {\vec {\mathbf r}}}\, \, $; ${\vec {\mathbf v}}$ is the velocity 
of the parton and $r\, = \, |{\vec {\mathbf r}}|$. The value of the
transverse dimension is taken as $R\sim 7$ fm. 

The quenched  spectrum convoluted with the transverse 
geometry of the partonic system can be written from (\ref{rate}) as
\begin{equation}
\frac{dN^{\rm{med}}}{d^2p_\bot}\, =
\, Q(p_\bot) \, \frac{dN^{\rm{vac}}}{d^2p_\bot}\, = \,
\frac{1}{2\pi^2 R^2} \ \int_0^{2\pi} \, d\phi \, \int_0^R \, d^2r \, \, 
\frac{dN (p_\bot+ \Delta E)}{d^2 p_\bot}\, \, \, . \label{supp}
\end{equation}

The $p_\bot$ distribution of charmed hadrons, $D$-mesons, produced in
hadron collisions were experimentally found~\cite{Alves} to be well
described by the following simple parameterization as
\begin{equation}
\frac{dN^{\rm{vac}}_H}{ d^2p_\bot}\, = \, C \left (\frac{1}{bM_C^2+p^2_\bot}
\right )^{n/2} \, \, \, , \label{hparam}
\end{equation}
where $b=1.4\pm0.3$, $n=10.0\pm1.2$ and $M_C=1.5$ GeV. 

The parameterization of the $p_\bot$ distribution exists 
in the literature~\cite{muel,Dokshitzer,dks} 
which describes the first RHIC light hadroproduction data for moderately
large values of $p_\bot$. In this case we consider the form given in 
Ref.~\cite{Dokshitzer} which reads as
\begin{equation}
\frac{dN^{\rm{vac}}_L}{ d^2p_\bot}\, = \, A \left (\frac{1}{p_0+p_\bot}
\right )^{m} \, \, \, , \label{lparam}
\end{equation}
where $m = 12.42$ and $p_0=1.71$ GeV/$c$.

The light hadron quenching using collisional energy loss rate~\cite{Thoma2}
was first anticipated~\cite{munshi0} to be of same order as that of
radiative~\cite{muel} ones, though
most of the previous studies insisted that the collisional energy loss
is insufficient to describe the medium modifications of hadron spectra.
We also here estimate the light hadron quenching 
(for details see Ref.~\cite{munshi0}) with the energy loss due to elastic 
collisions~\cite{Thoma2} 
and the energy loss rate averaged over parton species reads
\begin{equation}
-\, \frac{dE}{dL}\, = \, \frac{4}{3}\left(1+\frac{9}{4}\right )\,
\pi \alpha_s^2 T^2 \left ( 1+\frac{n_f}{6} \right ) \log \left [ 
2^{n_f/2(6+n_f)}\, 0.92 \, \frac{\sqrt{ET}}{m_g} \right ]\, . \label{pdedx}
\end{equation}

\begin{figure}
\begin{minipage}[h]{0.48\textwidth}
\centering{\includegraphics[height=1.2\textwidth,width=1.0\textwidth]{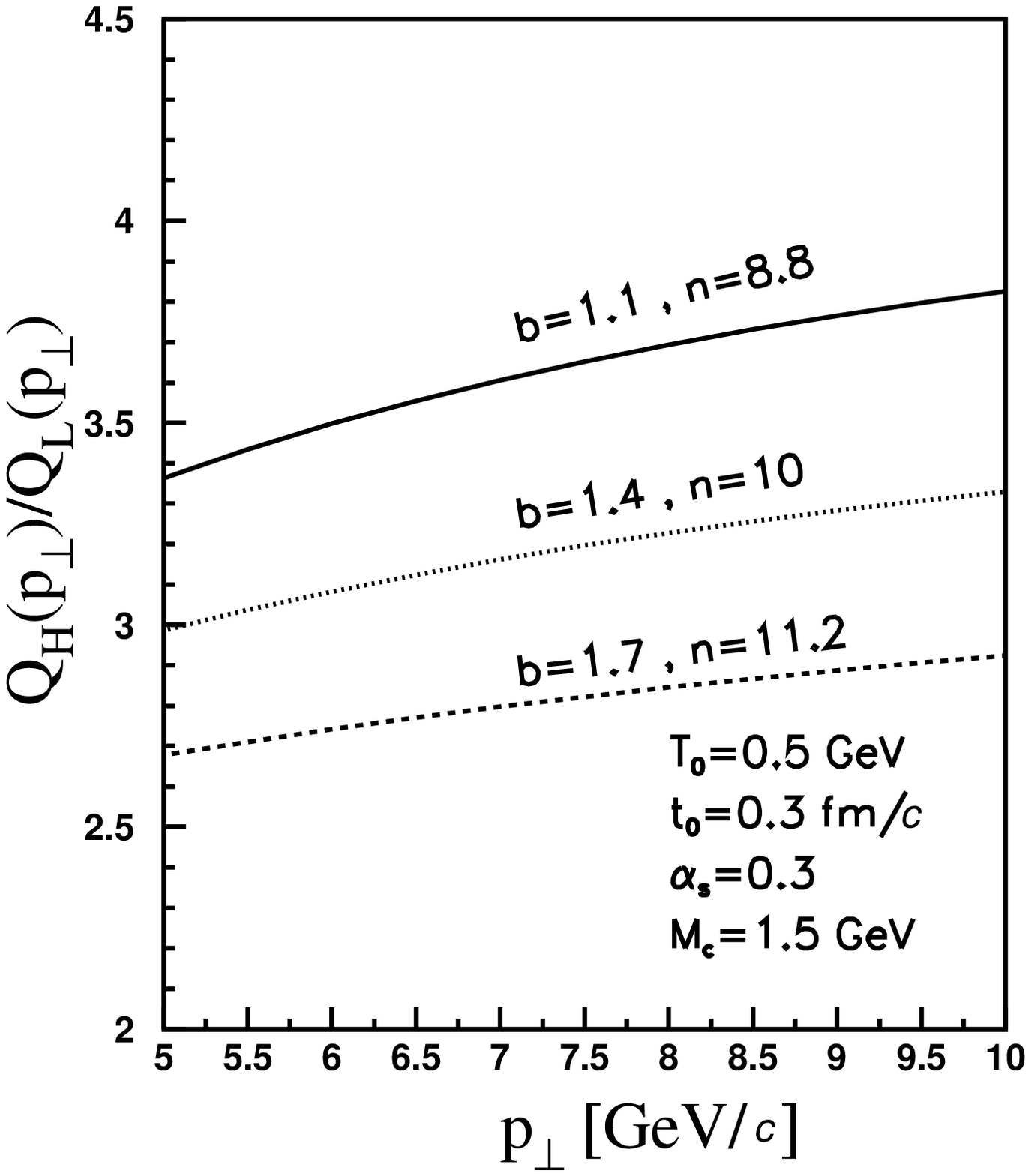}}
\end{minipage}
\begin{minipage}[h]{0.48\textwidth}
\centering{\includegraphics[height=1.2\textwidth,width=1.0\textwidth]{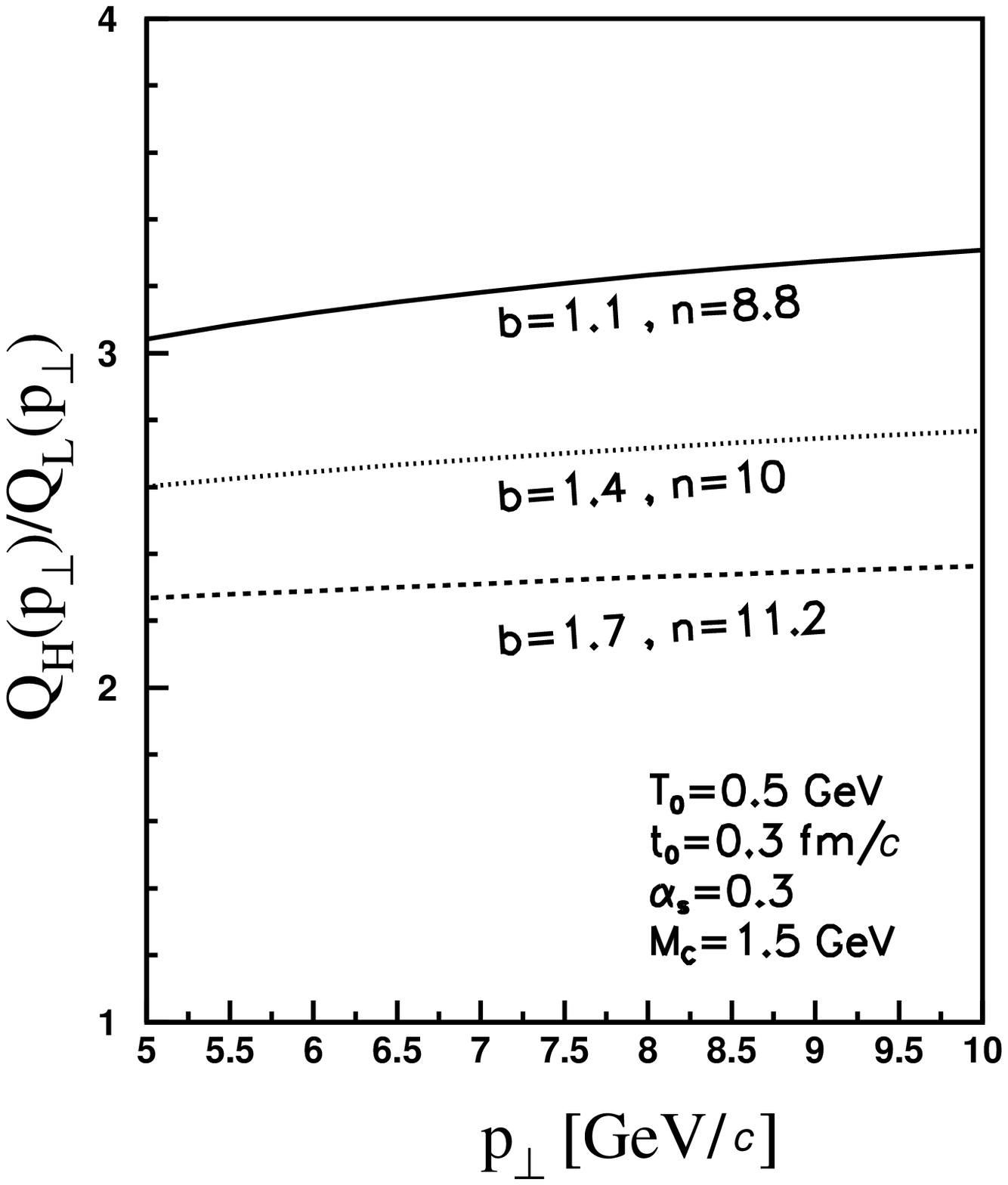}}
\end{minipage}
\vspace{-0.5in}
\caption{
Left panel: The ratio of charm to light quark quenching factors, 
$Q_H(p_\bot)/Q_L(p_\bot)$, as a function of 
transverse momentum $p_\bot$ with collisional energy loss.
Right pane: The ratio of charm to light quark quenching factors, 
$Q_H(p_\bot)/Q_L(p_\bot)$, as a function of 
transverse momentum $p_\bot$ using both collisional and radiative energy 
losses.
} 
\label{fig_ratio}
\end{figure}

We now illustrate in the left panel of Fig.~\ref{fig_ratio} the ratio 
of heavy to light quark quenching 
factors, ${Q_H(p_\bot)}/{Q_L(p_\bot)}$, as a function of transverse 
momentum $p_\bot$, using the collisional energy loss for plasma parameters
$T_0=0.5$ GeV, $t_0=0.3$ fm/$c$, $\alpha_s=0.3$ and the charm quark mass, 
$M_C=1.5$ GeV. As discussed earlier this ratio may reflect  heavy to light 
hadrons, $D/\pi$, ratio originating from the fragmentation of heavy and 
light quarks in heavy ion collisions. As shown the $D/\pi$ ratio is enhanced 
significantly as compared to $p-p$ collisions. 
The enhancement factor varies from $2.5$ to $4$ in the $p_\bot$ range 
(5-10) GeV/$c$, due to the uncertainties of different choices of parameter 
set to parameterize the heavy hadron spectra as depicted in (\ref{hparam}). 
However, the ratio also strongly depends on the quenching of light quark jets. 
The numerical estimate 
shows that the quenching of charm quarks is about a half that of light quarks. 
The light quarks, for a given $p_\bot$, lose 10$\%$ of their energy after 
traversing a distance 1 fm and around 40$\%$ after 5 fm~\cite{munshi0} whereas
the charm quark loses $5\%$ at 1 fm and around 20$\%$ at 5 fm (Fig.~\ref{fig_eloss_evo}). Because
of the large mass of the charm quark the $D$ meson will be formed in shorter
distance and hence charm quark would have less time to propagate in the medium
before transforming into the $D$ meson.
On the other hand, the light quarks would travel in the medium over a longer 
period and suffer more loss in energy than heavy quarks. The ratio is also 
found to be 
little more than that obtained earlier~\cite{Dokshitzer} considering only the
radiative energy loss with the appearance of the kinematical dead cone effect 
due to the 
finite heavy quark mass. This implies that the collision is one of the most
dominant mechanisms of energy loss in the medium.

As shown that both collisional and radiative energy losses are of same 
order in magnitude it would be interesting  to predict the $D/\pi$
ratio within our model considering radiative energy loss in addition to
collisional one. Since neither the drag and diffusion coefficients have
been calculated for other than collisional processes, it is not possible
to infer the impact of radiative processes directly within our model. 
In order to circumvent our lack of knowledge of radiative processes in terms
of the transport coefficients, we 
consider an alternative sets~\cite{Pol} within our model, obtained by 
multiplying the transport coefficients by a factor $K=2$. 
It is our hope that the experimental data will allow us to fix an approximate
value of $K$, if at all required.
In the right panel of Fig.~\ref{fig_ratio}, we plot the $D/\pi$ ratio for such 
a case with the same parameter 
set as before. 
As shown the ratio is little reduced compared to the collisional case and 
it varies from 2 to 3 within the $p_\bot$ range, (5-10) GeV/$c$. There is no 
striking change in the ratio mostly due to the cancellation of the introduced 
$K$ factor with reference to the collisional one for respective species.  
However, the small change can be attributed as the charm quark 
with the inclusion of radiative energy loss is relatively more quenched 
than that of the light quarks.

One may also add an interesting scenario, after the quark jet has hadronized
to leading particles, they would scatter with hadronic matter before 
decoupling. Considering $\sigma_{D\pi} << \sigma_{\pi\pi}$, {\it e.g.}, it is
most likely that the heavy mesons would decouple quickly from the hadronic 
phase. Pions would interact with each other via resonance formation and also 
with other light hadrons. This might lead to further enhancement of $D/\pi$ 
ratio. 

\section{Conclusion}
\label{sec_conl}

Apart from uncertainties in the various parameters describing the plasma 
and hadron spectra let us also have a look at some of the assumptions made 
in this work which may affect our findings. 
First, as discussed above, the momentum dependence of the drag and
diffusion coefficients, 
containing the 
dynamics of the elastic collisions, has been averaged out. A major advantage 
of this is the simplicity of the resulting differential equation. 
Of course, this simplification  
can lead to some amount of uncertainty in the quenching factor. Secondly, the 
entire discussion is based on the one dimensional Fokker-Planck equation and
the
Bjorken model of the nuclear collision, which may not be a very realistic 
description here but can provide a useful information of the 
problem.
However, extension to three dimension is indeed an ambitious goal,
which may cause that many of the considerations of the present work 
will have to be revised. 

In the present calculation
the hadron spectra for both light and heavy quarks have directly  been
used to calculate the quenching factor, $Q(p_\bot)$. This is equivalent
to assuming that a quark forms a hadron without much change in its energy.
In order to calculate 
the effects of the parton energy loss on the quenching pattern of high
$p_\bot$ partons in nuclear collisions, one should take into account the 
modification of the fragmentation function~\cite{Salgado,Wang} of a 
leading quark resulting from many soft interactions of the hard partons 
in the medium. This also causes a significant energy loss of parton 
prior to hadronization and changes the kinematic variables of the
fragmentation function~\cite{Wang}. This can also modify the quenching factor
and thus $D/\pi$ ratio.  

We show that the total collisional energy loss is almost same order as 
that of radiative energy loss for a static plasma. 
Considering the collisional energy loss
rates we obtain a momentum loss distribution for charm quarks by solving
the time evolution of Fokker-Plank equation for a thermally evolving plasma.
The total energy loss for an expanding plasma is found to be reduced by a
factor of 5 as compared to static case and does not depend linearly on the
system size. A ratio of heavy to light hadrons $D/\pi$ ratio is also estimated.
Though the collisions have different spectrum than radiation and
therefore contribute in a similar way to the suppression factor than 
anticipated earlier.

We now eagerly wait for the experimental data and
a detailed calculation have to be carried out before a realistic number
for the value of the $D/\pi$ ratio can be presented.
Nevertheless, the total collisional energy loss of a charm quark computed
within our simplified model may imply that the collision is one of the 
important energy loss mechanisms in the medium for the energy range (5-10) GeV
and this feature may be phenomenologically important. 
Also, the results for $D/\pi$ presented within this model 
is not definitive but can provide a very intuitive picture of the medium energy 
loss for partons in moderately large $p_\bot$. 


\section*{Acknowledgment} 
The author is thankful to  
 J. Alam, S. Raha, D. K. Srivastava and M. H. Thoma for various useful 
discussion.


\begin{thebibliography}{99}
\bibitem{Pluemer} M. Gyulassy and M. Pl\"umer, Phys. Lett. B {\bf 243} (1990)
432.
\bibitem{RHIC} K. Adcox et al. [PHENIX Collaboration], Phys. Rev. Lett. 
{\bf 88} (2002) 022301; 
C. Adler et al. [STAR Collaboration], Phys. Rev. Lett. {\bf 89} (2002) 202301; 
{\it ibid} {\bf 90} (2003) 032301; {\it ibid} {\bf 90} (2003) 082302; 
S. S. Adler et. al.[PHENIX Collaboration], Phys. Rev. Lett {\bf 91} (2003) 
072301; 
 B.B. Back et al. [PHOBOS Collaboration], {\it nucl-ex/0302015}.
\bibitem{Gyulassy} M. Gyulassy, P. Levai, and I. Vitev, Phys. Lett. B {\bf 538} 
(2002) 282; M. Gyulassy, I. Vitev, X. N. Wang, Phys. Rev. Lett. {\bf 86}
 (2001) 2537.
\bibitem{Salgado} C. A. Salgado and U. A. Wiedemann, Phys. Rev. Lett. {\bf 89}
 (2002) 092303.
\bibitem{Wang}E. Wang and X. N. Wang, Phys. Rev. Lett. {\bf 89} (2002) 162301
\bibitem{Baier}R. Baier, Y. L. Dokshitzer, A. H. Mueller, and D. Schiff,
JHEP {\bf 0109} (2001) 033.
\bibitem{muel} B. M\"uller, Phys. Rev. C {\bf 67} (2003) 061901(R).
\bibitem{munshi0} M. G. Mustafa and M. H. Thoma; {\it hep-ph/0311168}, 
Acta Phys. Hung. A {\bf 22}/1-2 (2005) 93, Edward Teller Memorial Volume, 2004.
\bibitem{Light}
B. B. Back et al.[PHOBOS Collaboration],  Phys. Rev. Lett. {\bf 91} (2003) 
072302;
S. S. Adler et al. [PHENIX Collaboration], Phys. Rev. Lett. {\bf 91} (2003) 
072303;
J. Adams et al.[STAR Collaboration],  Phys. Rev. Lett. {\bf 91} (2003) 072304;
I. Arsene et al.[BRAHMS Collaboration],  Phys. Rev. Lett. {\bf 91} (2003) 
072305;
\bibitem{Heavy} K. Adcox et al. [PHENIX Collaboration], Phys. Rev. Lett. 
{\bf 88} (2002) 192303.
\bibitem{Bjorken} J.D. Bjorken, Fermilab preprint {\bf 82/59-THY} (1982, unpublished).
\bibitem{svet} B. Svetitsky, Phys. Rev. D {\bf 37} (1988) 2484.
\bibitem{Thoma1} M.H. Thoma and M. Gyulassy, Nucl. Phys. B {\bf 351} (1991)
491.
\bibitem{Mrowczynski} S. Mr\'owczy\'nski, Phys. Lett. B {\bf 269} (1991) 383.
\bibitem{Koike} Y. Koike and T. Matsui, Phys. Rev. D {\bf 45} (1992) 3237.
\bibitem{Abhee} A. K. Dutta-Mazumder, J. Alam, P. Roy and B. Sinha, 
{\it hep-ph/0411015} (To appear in Phys. Rev. C). 
\bibitem{munshi} M. G. Mustafa, D. Pal and D. K. Srivastava, Phys. Rev. 
C {\bf 57} (1998) 889.
\bibitem{Hendrik} H. van Hees and R. Rapp, {\it nucl-th/0412015} 
(Phys. Rev. C in press).
\bibitem{BP} E. Braaten and R.D. Pisarski, Nucl. Phys. B {\bf 337} (1990) 569.
\bibitem{Braaten1} E. Braaten and M.H. Thoma, Phys. Rev. D {\bf 44} (1991)
1298.
\bibitem{Braaten2} E. Braaten and M.H. Thoma, Phys. Rev. D {\bf 44} (1991) 
2625(R).
\bibitem{Vija} H. Vija and M.H. Thoma, Phys. Lett. B {\bf 342} (1995) 212.
\bibitem{Romatschke} P. Romatschke and M. Strickland, Phys. Rev. D {\bf 69}
(2004)065005;  P. Romatschke and M. Strickland,  {\it hep-ph/0308275}.
\bibitem{Moore} G. D. Moore and D. Teaney, {\it hep-ph/0412346}.
\bibitem{Thoma2} M. H. Thoma, Phys. Lett. B {\bf 273} (1991) 128.
\bibitem{Thoma3} M. H. Thoma, J. Phys. G {\bf 26} (2000) 1507.
\bibitem{Thoma4} M. H. Thoma, Eur. Phys. J. C {\bf 16} (2000) 513.
\bibitem{BSZ} R. Baier, D. Schiff, and B.G. Zakharov, Ann. Rev. Nucl. Part. 
Sci. {\bf 50} (2000) 37.
\bibitem{WW} E. Wang and X.-N. Wang, Phys. Rev. Lett. {\bf 87} (2001) 142301.
\bibitem{munshi1}M. G. Mustafa, D. Pal, D. K. Srivastava, and M. H. Thoma,
Phys. Lett. B {\bf 428} (1998) 234.
\bibitem{Vogt} Z. Lin, R. Vogt and X.N. Wang, Phys. Rev. C {\bf 57} (1998) 899;
Z. Lin and R. Vogt, Nucl. Phys. B {\bf 544} (1999) 339.
\bibitem{Shuryak} E. Shuryak, Phys. Rev. C {\bf 55} (1997) 961.
\bibitem{Dokshitzer} Y. L. Dokshitzer and D. E. Kharzeev, Phys. Lett. B 
{\bf 519} (2001) 199.
\bibitem{Djordjevic} M. Djordjevic and M. Gyulassy, Phys. Lett. B {\bf 560}
 (2003) 37.
\bibitem{Djordjevic1} M. Djordjevic and M. Gyulassy, Nucl. Phys. A {\bf 733} 
(2004) 265.
\bibitem{Armesto} N. Armesto, C. A. Salgado and U. A. Wiedemann, Phys. Rev. D
{\bf 69} (2004) 11403.
\bibitem{opacity} M. Gyulassy, P. Levai, and I. Vitev, Nucl. Phys. B 
{\bf 594} (2001) 371.
\bibitem{Quad} U. A. Wiedemann, Nucl. Phys. B {\bf 558} (2000) 303, 
{\it ibid} {\bf 582} (2000) 409.
\bibitem{bales} R. Balescu, {\it Equilibrium and Non-Equilibrium Statistical
Mechanics} (Wiley, New York, 1975).
\bibitem{somenath} S. Chakrabarty and D. Syam, Lett. Nuovo Cimento {\bf 41}
(1984) 381.
\bibitem{rafelski} D. B. Walton and J. Rafelski, Phys. Rev. Lett. {\bf 84}
(2000) 31.
\bibitem{jane} J. Alam, S. Raha, and B. Sinha, Phys. Rev. Lett. {\bf 73}
(1994) 1895; P. Roy, J. Alam, S. Sarkar, B. Sinha, and S. Raha, 
Nucl. Phys. A {\bf 624} (1997) 687.
\bibitem{bjor} J. D. Bjorken, Phys. Rev. D {\bf 27} (1983) 140.
\bibitem{william} W. E. Williams, {\it Partial Differential Equations} 
(Clarendon Press, Oxford, 1980).
\bibitem{Wang1} X. N. Wang, Phys. Rev. Lett. {\bf 81} (1998) 2655.
\bibitem{Alves} C. A. Alves {\it et al.} [E796 Collaboration], Phys. Rev. 
Lett. {\bf 77} (1996) 2392.
\bibitem{dks} R. J. Fries, B. M\"uller and D. K. Srivastava, Phys. Rev. Lett.
{\bf 90} (2003) 132301.
\bibitem{Pol} P. B. Gossiaux, V. Guiho, and J. Aichelin, {\it hep-ph/0411324}.
\end{thebibliography}
\end{document}